\documentclass[aps,pre,letterpaper,twocolumn,groupedaddress,showpacs,]{revtex4-1} 

\usepackage{amsmath,amssymb,graphicx}
\usepackage{color}
\usepackage{natbib}

\begin{document}

\title{Structured scale-dependence in the Lyapunov exponent of a Boolean chaotic map}

\author{Seth D. Cohen}

\affiliation{Miltec Corp., A Ducommun Company, 678 Discovery Drive, Huntsville, Alabama 35806  USA}

\date{\today}

\begin{abstract} We report on structures in a scale-dependent Lyapunov exponent of an experimental chaotic map that arise due to discontinuities in the map. The chaos is realized in an autonomous Boolean network which is constructed using asynchronous logic gates to form a map operator that outputs an unclocked pulse-train of varying widths. The map operator executes pulse-width stretching and folding and the operator's output is fed back to its input to continuously iterate the map. Using a simple model, we show that the structured scale-dependence in the system's Lyapunov exponent is the result of the discrete logic elements in the map operator's stretching function.

\end{abstract}

\maketitle

\section{Introduction}

Understanding the distinct roles of noise and determinism is important for all experimental chaotic systems. We examine a scale-dependent Lyapunov exponent (SDLE) of a Boolean chaotic system that iterates the dynamics a one-dimensional (1D) map. A SDLE was studied previously as a way to distinguish the entropy of microscopic noise from the entropy of macroscopic chaos \cite{Gao1999,Cencini2000,Gao2006}. This distinction and the transition from a microscale entropy to a macroscale entropy is of increasing importance as physical random number generators that use chaotic systems as entropy sources continue to be developed \cite{Uchida2008, Reidler2009, Argyris2010, Sunada2012, RosinRandom2013}. Our experiment iterates a macroscopic tent map with microscopic noise and an intermediate scaling caused by mesoscopic discontinuities in the map. Here, we demonstrate experimentally that the scaling of these discontinuities manifests as a structured scale-dependence in the Lyapunov exponent of our Boolean chaotic system.

Boolean chaos is a term used to describe the phenomenon of deterministic dynamics in unclocked Boolean networks with an exponential divergence of neighboring trajectories. Originally, theories of continuous, ideal Boolean networks predicted non-repeating switching in certain networks, but without chaos \cite{Ghil1984}. Contrary to this prediction, recent experiments showed that physical logic gates can introduce non-ideal effects that give rise to such chaos \cite{Zhang2009}. Boolean chaos has been reported in both autonomous and driven networks \cite{Zhang2009, Blakely2014}, both of which yield complex, multi-dimensional dynamics. One-dimensional (1D) Boolean chaotic maps were theorized for specific non-ideal effects \cite{Cavalcante2010}. Here, we examine the first experimental 1D Boolean chaotic map and report its unique, multi-scaled features.

Our experimental setup is influenced by recent studies of non-chaotic autonomous Boolean systems. In particular, asynchronous networks of logic gates were studied as excitable systems with synchronization patterns \cite{RosinExcitability2012,RosinPattern2013} and phase oscillators with chimera states \cite{RosinPhase2014, RosinChimera2014}. One appealing feature of these Boolean systems is that they can be implemented entirely on a field-programmable gate array (FPGA), a common component in modern electronics. This platform allows for large dynamical networks of asynchronous logic gates to easily be built \cite{RosinExcitability2012,RosinRandom2013,RosinPattern2013,RosinPhase2014,RosinChimera2014}. The implementation of our experiment is also facilitated by an FPGA, where we note that our system is not a simulation on a finite-state-machine (see \cite{Tanougast2011} and references therein); it is an unclocked system with non-zero entropy and a potential continuum of dynamical states that are subject to analog effects and experimental noise.

\begin{figure}[b!]
\resizebox{8.0cm}{!}{\includegraphics{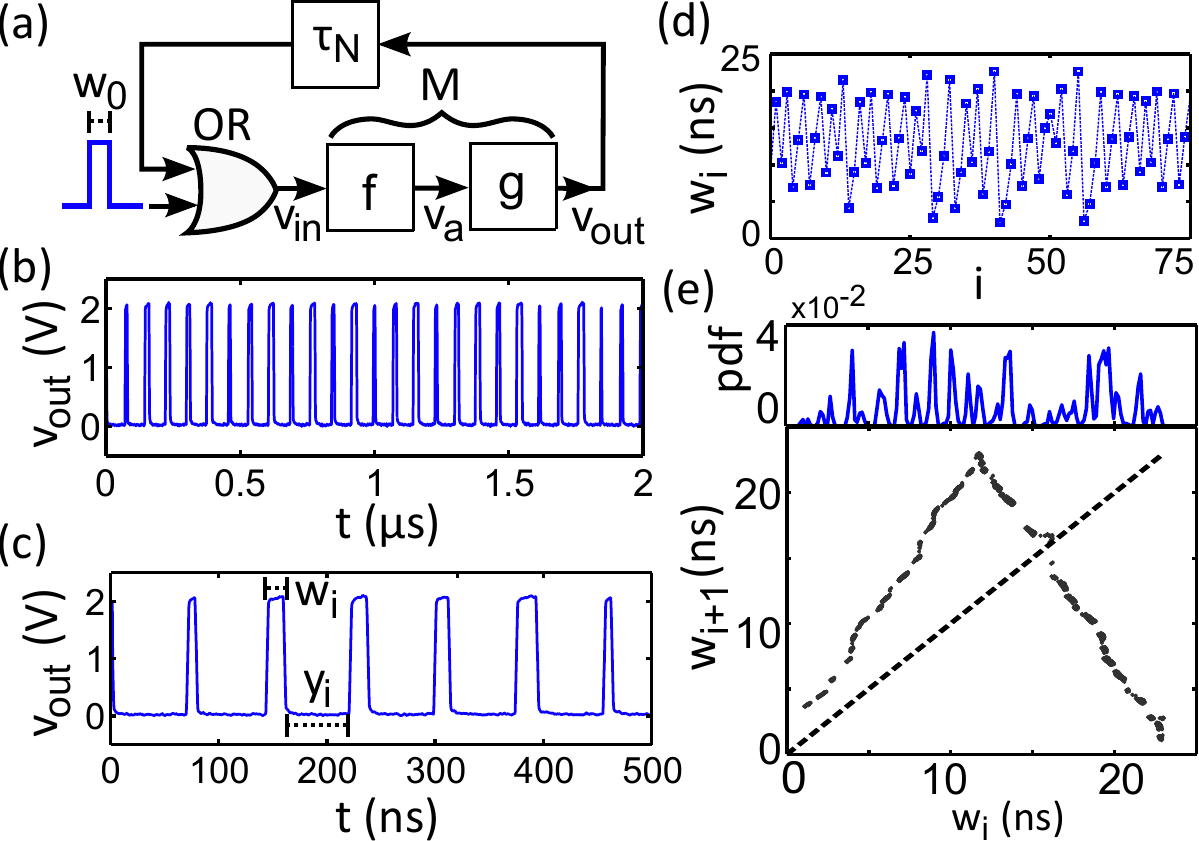}}
\caption{ (a) Experimental system with operator $M$, initial pulse of width $w_0$ injected via OR gate, input voltage $v_{\text{in}}$, auxiliary voltage $v_{\text{a}}$, output voltage $v_{\text{out}}$, and feedback loop of delay $\tau_N \sim 60$ ns (realized using a $N = 200$ NOT gates) on an Altera Cyclone IV (EP4CE115F29C7N) using $\sim 1\%$ FPGA resources. (b)-(c) $v_{\text{out}}$ from $w_0$ = 16 ns. (d) $w_i$ as a function of iteration $i$. (e) Return map $(w_i,w_{i+1})$ with dashed line of slope 1 and normalized histogram of states showing the probability distribution function (pdf) of $w_i$.}
\label{fig:Dynamics}
\end{figure}

\section{Experimental System}

Our experimental system is shown in Fig. \ref{fig:Dynamics}a, which is initialized by an input voltage pulse of initial pulse-width $w_0$. This pulse drives a map operator $M$ which consists of two separate functions: a pulse-width folding-function $f$ and a pulse-width gain-function $g$ that approximately doubles a pulse's width (details provided later), where this combination of folding and stretching are sufficient conditions to see chaos \cite{Strogatz}. The output voltage of $M$ is labeled as $v_{\text{out}}$, and a delay line routes $v_{\text{out}}$  back to the input of $M$, where the delay is long enough to ensure only one pulse is in this feedback loop at a time. We note that this is a time-delay system with an infinite dimensional phase space, but neighboring pulses do not interact, allowing for 1D dynamics. This system remains in a stable steady state $v_{\text{out}}=0$ V until we inject a pulse, and thus $w_0$ serves as its initial condition.

After an initial pulse is injected, the system produces a self-sustaining pulse-train. In Figs. \ref{fig:Dynamics}b-c, we plot the temporal evolution of $v_{\text{out}}$, which contains pulses that occur with non-repeating pulse-widths $w_{i}$ and non-repeating spacings $y_{i}$. Thus, the transition times in $v_{\text{out}}$ are the state variables of the chaos. This is different from analog chaotic circuits that iterate 1D maps in discrete or continuous time and use voltage or current as the state variables \cite{Morie2000,Corron2010}. The power spectral density of $v_{\text{out}}$ (not shown) is broadband with prominent frequency components at integer multiples of $1/T \sim 13 \text{ MHz}$, where $T = \bar{w} + \bar{y} \sim 76$ ns is the average pulse-repetition period and can be adjusted with the feedback delay.

To analyze the dynamics, we study $w_i$, which is plotted in Fig. \ref{fig:Dynamics}d; as we will show later, $y_i$ is a function of $w_i$ and contains no new information about the dynamics. We construct the return map using $(w_{i},w_{i+1})$ in Fig. \ref{fig:Dynamics}e, which shows a 1D structure similar to a tent map \cite{Strogatz}. Figure \ref{fig:Dynamics}e also shows that the density of the return map is non-uniform, which differs from an ideal tent map.

We fit Fig. \ref{fig:Dynamics}e using a piecewise-linear function
 \begin{equation}
f(w_{\text{i}}) = w_{i+1} \sim \left\{
        \begin{array}{ll}
            m w_i, & w_i \leq \tau_n \\
            m(2\tau_{n}- w_{i}), &  \tau_n< w_{i} \leq 2\tau_n
        \end{array}
    \right.
\label{eq:tentfold}
\end{equation}
where $m$ is the average slope and $\tau_n$ is the folding point. The fit yields $m_{\text{fit}}=1.95 \pm 0.01$ and $\tau_{n,\text{fit}}=11.7 \pm 0.01$ ns, which shows that the map is not a tent-map of full height. We note that a tent map of full height in the experimental Boolean implementation can only show transient chaos before collapsing to the steady state $v_{\text{out}}=0$ V due to short-pulses rejection (SPR) by the physical logic gates \cite{Zhang2009}. The slope $m=1.95$ limits the grammar of the map and prevents pulse widths $w_i \lesssim 1$ ns, allowing for non-transient chaos. 

\subsection{Lyapunov Exponent Calculations}

Interestingly, we might approximate the Lyapunov exponent $\lambda$ using $m_{\text{fit}}$ such that $\lambda \sim \text{ln}(1.95)/T$ \cite{Strogatz}. However, using this fit of the return map to estimate $\lambda$ assumes a continuous map. As we will discuss later, the discrete logic gates in the system's design create discontinuities in the output of the operator $M$. To avoid assumptions about the experimental system, we instead compute a SDLE from $w_i$.

\begin{figure}[t!]
\resizebox{8.25 cm}{!}{\includegraphics{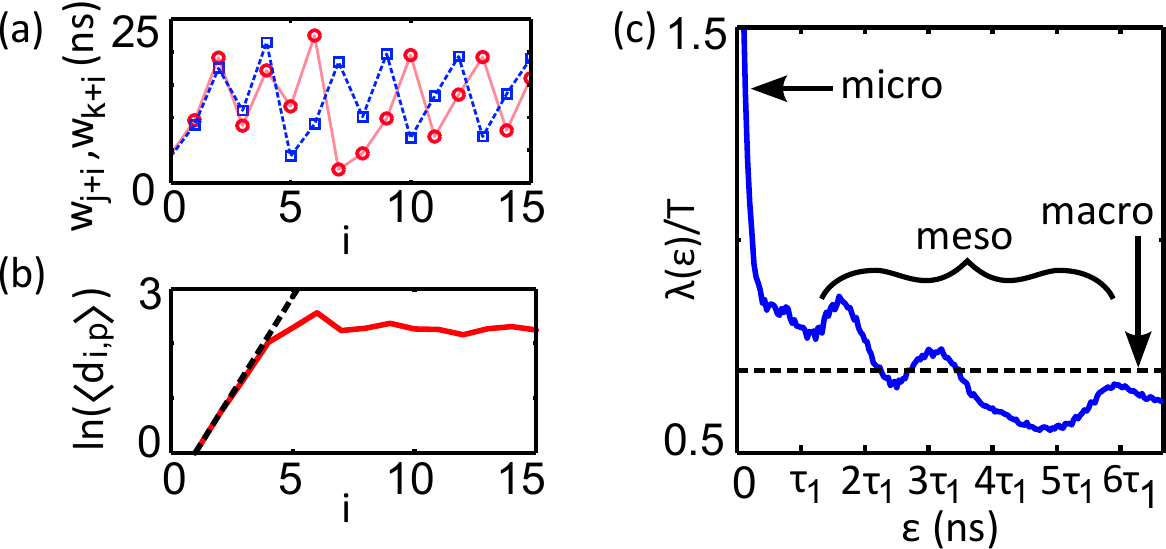}}
\caption{ (a) $w_{j+i}$ (dashed line) and $w_{k+i}$ (solid line) that diverge after several iterations. (b) Averaged distances $\text{ln}(\langle d_{i,p}\rangle)$ for $\epsilon=700$ ps and $\Delta \epsilon= 5$ ps, with fitted line of slope $\sim \text{ln}(2)$. (c)  $\lambda(\epsilon)/T$ with a dashed line to indicate $\lambda(\epsilon) /T = \text{ln}(2)$. In the figure, we label the different scalings of the SDLE that can be attributed to the microscopic noise, mesoscopic features, and the macroscopic chaos.}
\label{fig:lyapunov}
\end{figure}

We define a SDLE $\lambda (\epsilon)$ as the divergence of neighboring trajectories along the return map, where neighbors $w_j$ and $w_k$ satisfy 
 \begin{equation}
\epsilon <|w_j-w_k| < \epsilon+\Delta\epsilon,
\label{eq:epsil}
 \end{equation}
where $\epsilon$ determines the scale of $\lambda (\epsilon)$ and $\Delta \epsilon$ is a fixed width that determines the window size to search for neighboring points. For a given $\epsilon$, we find all $w_j$ and $w_k$ that satisfy Eq. [\ref{eq:epsil}] and define the initial trajectory separation as $d_{0,p} =|w_j-w_k|$, where $p$ is the index of $P$ total neighboring pairs. Next, we follow the trajectories of each neighboring pair $(w_j, w_k)$ and calculate their normalized divergences $d_{i,p}=|w_{j+i} -w_{k+i}|/d_0$ for $i=1$ to $i=15$. An example of two neighboring trajectories $w_{j+i}$ and $w_{k+i}$ is shown in Fig. \ref{fig:lyapunov}a. We then average $d_{i,p}$ over all $p$. An example of $\langle d_{i,p}\rangle$ for a given $\epsilon$ shown in Fig. \ref{fig:lyapunov}b on a log scale. 

Lastly, we use a linear fit of $\langle d_{i,p}\rangle$ (up to $i = 2$), where the slope of the linear fit gives the initial exponential separation of the neighboring trajectories for a given $\epsilon$. We define this initial exponential separation as $\lambda(\epsilon)$. The example in Fig. \ref{fig:lyapunov}b yields $\lambda(\epsilon)/T \sim 0.7$, where we divide by $T$ to provide a unitless quantity. Following this procedure, we calculate $\lambda(\epsilon)$ for a scan of $\epsilon$ in 10 ps steps and the result is shown in Fig. \ref{fig:lyapunov}c. In the figure, we use a scaling reference of $\tau_1$, which is the approximate delay time through a single logic element. We note that $\lambda(\epsilon)/T > 0$ is an indicator of trajectory separation at an exponential rate.

Figure \ref{fig:lyapunov}c demonstrates that the experimental system exhibits exponential separation from both noise and chaos. Neighboring points on the return map with $d_o < \tau_1$ have an initial separation that is dominated by microscopic noise, while for $d_o > \tau_1$ the SDLE shows oscillations in the divergence rate at frequency $\sim 1/(2\tau_1)$ near $\lambda(\epsilon)/T \sim \text{ln}(2)$ (the value of the Lyapunov exponent for a continuous tent-map of the full height). Thus, the rate of divergence shows a structured scale-dependence about a global rate that is described by the macroscopic features of the return map. Studying this phenomenon is important for understanding the transition from microscopic noise to macroscopic chaos. In the remainder of this paper, we briefly outline the design of Fig. \ref{fig:Dynamics}a and examine the map operator's functions to motivate a simple model that exhibits a similarly structured SDLE.

\subsection{Circuit Designs}

Our experimental system exploits the propagation delays of pulses as they transmit through logic gates. In the simplest example, the feedback loop $\tau_N$ in Fig. \ref{fig:Dynamics}a acts as a continuous delay line that routes pulses from the output of the map operator back to its input. This delay line is constructed using cascaded NOT gates \cite{RosinExcitability2012}, where even numbers of NOT gates are used to reduce asymmetries between rise and fall times of pulse edges that propagate \cite{Zhang2009} and preserve pulse widths. The number of NOT gates $n$ sets the propagation delay $\tau_{n}$.

\begin{figure}[t!]
\resizebox{8.0cm}{!}{\includegraphics{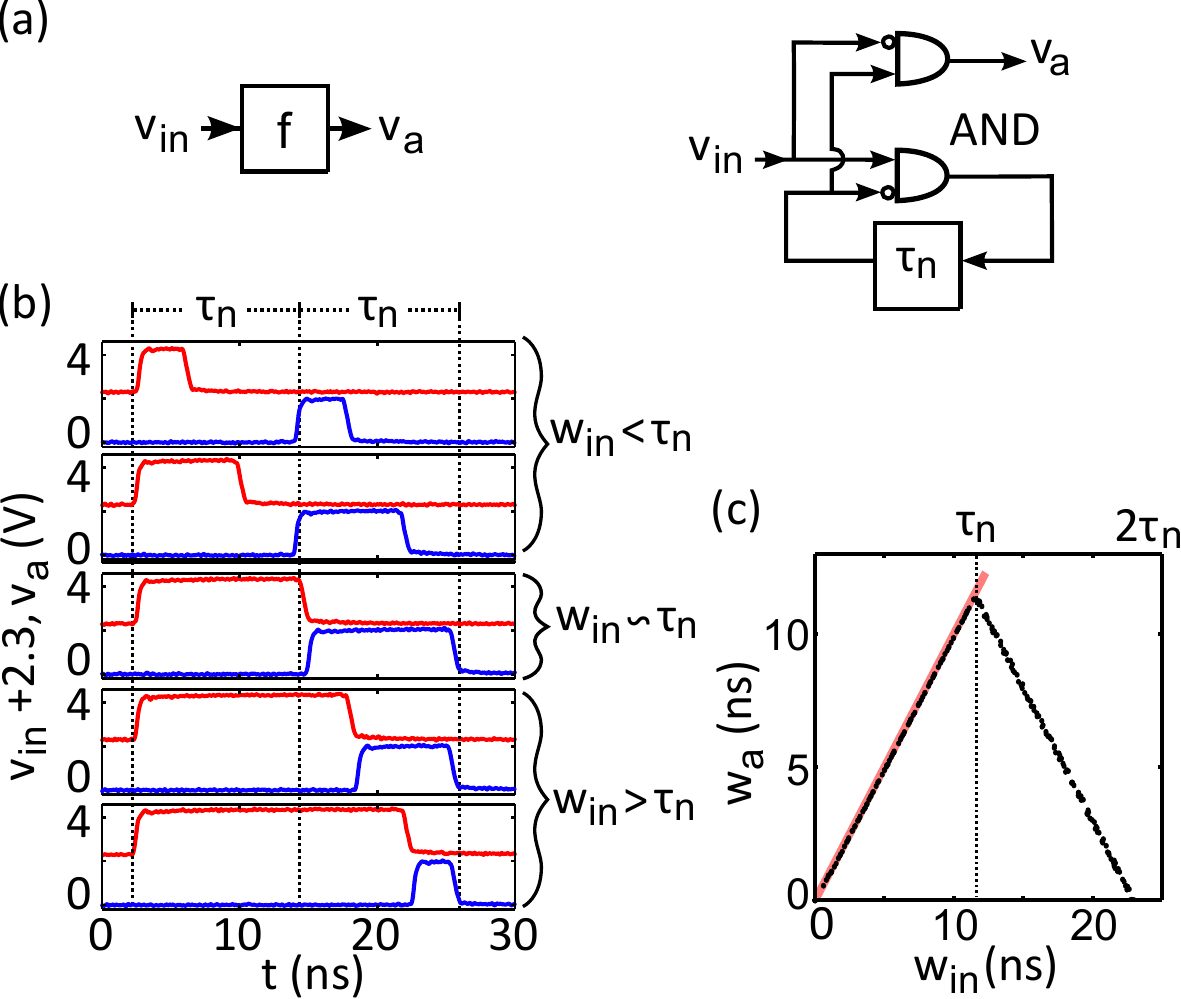}}
\caption{Pulse-width folding circuit. (a) $f$ with $\tau_{n}$ for $n=40$ and AND gates to map $v_{\text{in}}$ to $v_{\text{a}}$. (b) $v_{\text{in}}$ (top) and $v_{\text{a}}$ (bottom) using a Tektronix TDS7254B digital oscilloscope, Hewley Packard Pulse Generator (PG) 8116A, and the FPGA. A buffer gate at the input of the FPGA removes analog characteristics from the PG. (d) $w_{\text{a}}$ for values of $w_{\text{in}}$ in 100 ps steps and dashed line of slope 1. We note $f$ repeats every $2\tau_n$.}
\label{fig:TentMapFolding}
\end{figure}

In the map operator, a folding function $f$ is implemented with the circuit in Fig. \ref{fig:TentMapFolding}a. In the figure, $v_{\text{in}}$ and $v_{\text{a}}$ are signals for input pulses of width $w_{\text{in}}$ and output pulses of width $w_{\text{a}}$, respectively, such that $w_{\text{a}} \sim w_{\text{in}}$ for $w_{\text{in}} \leq \tau_n$ and $w_{\text{a}} \sim (2\tau_{n}-w_{\text{in}})$ for $\tau_n< w_{\text{in}} \leq 2\tau_n$. To illustrate this circuit's folding, in Fig. \ref{fig:TentMapFolding}b we plot experimental examples of $(v_{\text{in}}, v_{\text{a}})$ of the folding circuit, and in Fig. \ref{fig:TentMapFolding}c, we scan $w_{\text{in}}$ and plot the respective $w_{\text{a}}$. We model $w_{\text{a}} = f(w_{\text{in}})$ from Eq. [\ref{eq:tentfold}] with $m = 1$, and we note $f(w_{in})$ does not address $y_i$, but based on the folding circuit's operation, we derive $y_i = \tau_N + \tau_n - w_{i}$ for $w_{i} \leq \tau_n$ and $y_i = \tau_N$ otherwise.

\begin{figure}[t!]
\resizebox{8.5cm}{!}{\includegraphics{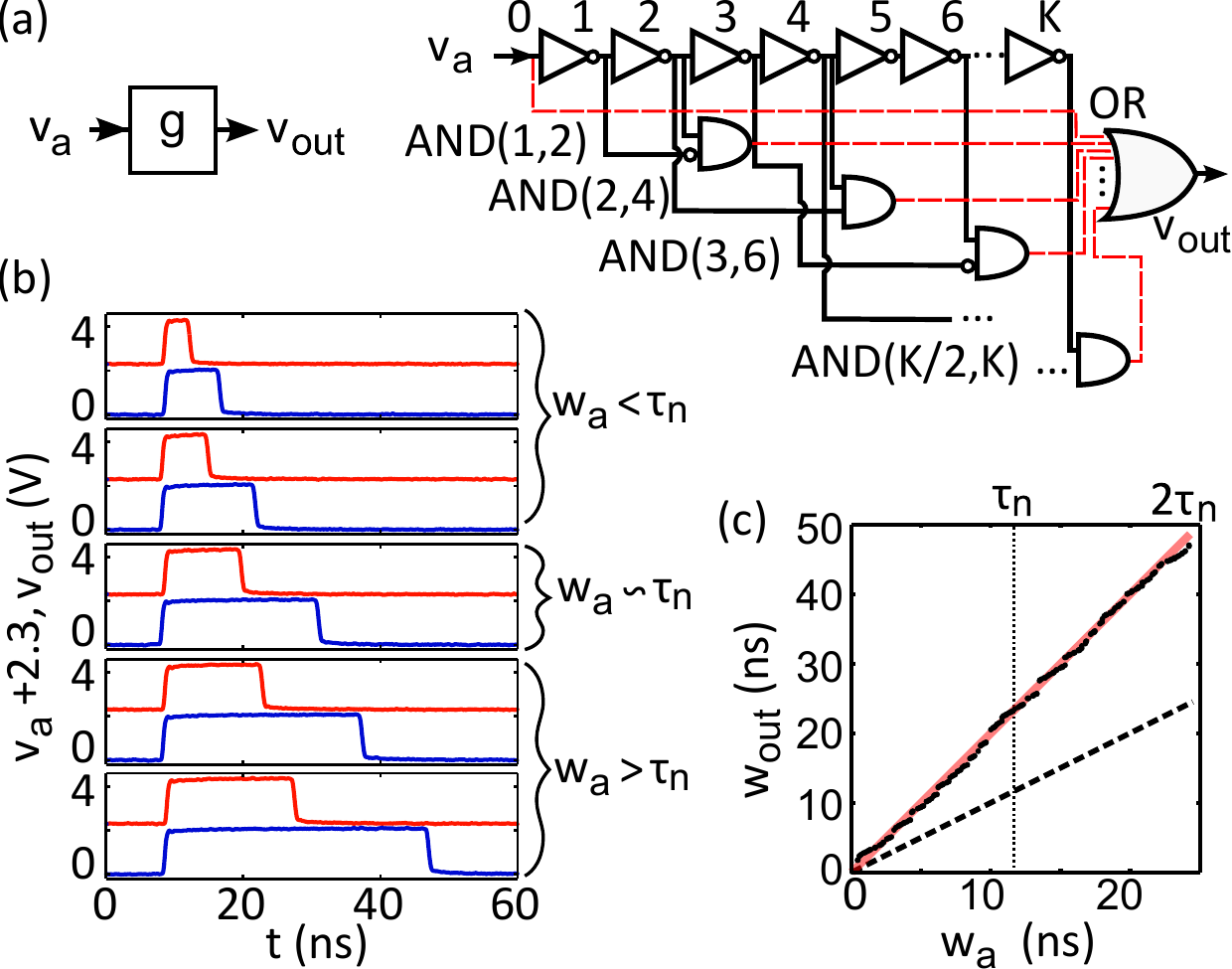}}
\caption{Pulse-width gain circuit. (a) $g$ with ($v_{\text{a}},v_{\text{out}}$), $K=200$ NOT gates, and $K/2$ AND gates that drive a $K/2$-input OR gate, realized by a tree of gates with balanced path-lengths. Odd-numbered AND gates have one input inverted to balance the logic. (b) $v_{\text{a}}$ (top) and $v_{\text{out}}$ (bottom). (c)  $w_{\text{out}}$ for $w_{\text{a}}$ at 100 ps steps and dashed (solid) line of slope 1 (2).}
\label{fig:Stretching}
\end{figure}

The pulse-width gain function $g$ of the map operator is shown Fig. \ref{fig:Stretching}a. In the figure, an input pulse $v_{\text{a}}$ is launched into a delay line of NOT gates, where AND gates compare the outputs of gates $(k,2 k)$, where $k$ is the index number of $K$ total NOT gates such that $2 k_{\text{max}} = K$. The AND-gate outputs drive a multi-input OR-gate, which outputs a pulse $v_{\text{out}}$ of width $w_{\text{out}}$ for $\tau_K > w_{\text{out}}$, where $\tau_K$ is the delay through $K$ gates. To illustrate the pulse-width gain from this circuit, we plot examples of $(v_{\text{a}},v_{\text{out}})$ and a scan of $(w_{\text{a}},w_{\text{out}})$ in Figs. \ref{fig:Stretching}b-c, respectively. The resulting waveforms show approximate pulse-width doubling and the characterization of $(w_{\text{a}},w_{\text{out}})$ has average slope $\sim 2$.

\section{Numerical Simulations}

However, the discrete nature of the AND-gate comparisons of the delay line in Fig. \ref{fig:Stretching}a creates regularly-spaced discontinuities that are not resolved in Fig. \ref{fig:Stretching}c. Based on these discrete comparisons, we model the discontinuous pulse-width gain as
\begin{equation}
 w_{\text{out}} = g(w_{\text{a}}) \sim 2 \tau_{1} \lfloor w_{\text{a}}/\tau_{1}\rfloor+ h(w_{\text{a}}-\tau_1\lfloor w_{\text{a}}/\tau_{1}\rfloor),
\label{eq:discont}
\end{equation}
where $\tau_{1} \lfloor w_{\text{a}}/\tau_{1}\rfloor$ is a measure of $w_{\text{a}}$ in single gate-delays, and $h$ is a function that describes the width of an output pulse for a single gate. As $w_{\text{in}}$ increases in Eq. [\ref{eq:discont}], $g$ is discontinuous and increases by steps of $\tau_1$. We define $h(0)=0$ such that, when $w_{\text{in}}$ is an integer multiple of $\tau_1$ ($w_{\text{a}}-\tau_1\lfloor w_{\text{a}}/\tau_{1}\rfloor=0)$, the pulse-width gain is exactly 2. When $w_{\text{a}}$ is not an integer multiple of $\tau_1$, $h$ provides a corrective term that describes the continuous growth of pulse widths, where the input to $h$ resets at each multiple of $\tau_1$. Thus, the function $g$ is has an average slope $\sim2$ with discontinuities spaced regularly by $\tau_1$ and local slope(s) $h'(w_{\text{in}})$ in between each discontinuity.

For simplicity, we let $h(w_{\text{a}})=w_{\text{a}}$ such that the map $w_{i+1}=M(w_{i})=g(f(w_{i}))$ is an example of a piecewise-linear system that exhibits noise-induced chaos. Noise-induced chaos occurs in chaotic systems that only show periodic or steady state dynamics without the presence of noise \cite{Gao1999}. Similar models with discontinuities have been previously studied with the use of a SDLE, where with enough noise, these simulated chaotic maps exhibit characteristics of their macroscopic map structures \cite{Cencini2000}. A different choice for $h$, such as a nonlinear function, can yield chaos without noise, but we choose the simplest chaotic model with mesoscopic discontinuities and microscopic noise. Noise in the FPGA causes jitter in $w_i$, where we measure the jitter to be approximately Gaussian with standard deviation (STD) $\sigma \sim 90$ ps. We simulate the map $w_{i+1}=M(w_{i})$ using Eqs. [\ref{eq:tentfold}-\ref{eq:discont}] with $m=1$, $\tau_n = 12$ ns, $\tau_1 = 0.3$ ns, and additive-white-Gaussian-noise at every iteration (STD $=\sigma$).

\begin{figure}[t!]
\resizebox{8cm}{!}{\includegraphics{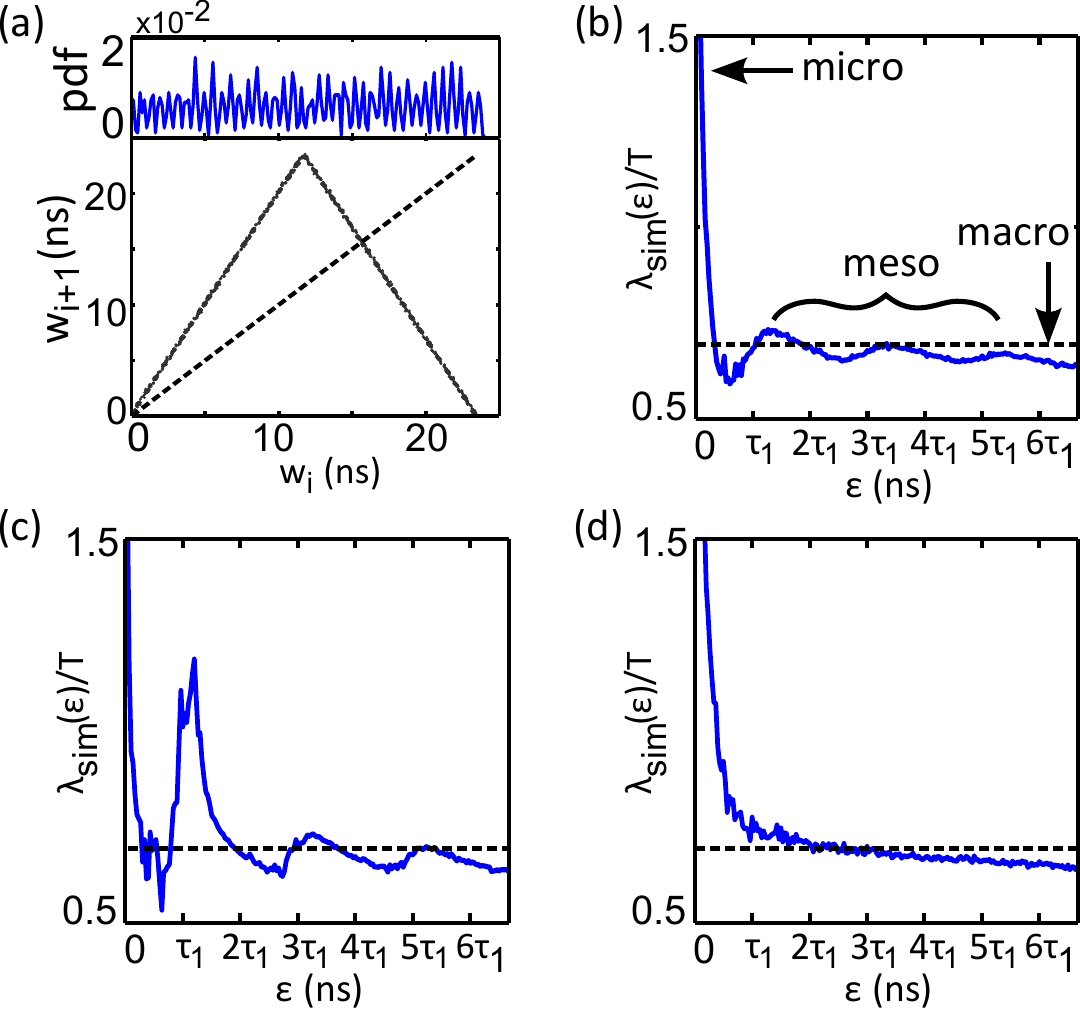}}
\caption{(a) Simulated return map $(w_i,w_{i+1})$ and pdf from the map $w_{i+1}=M(w_i)$ for $i = 1$ to $4000$ with noise level $\sigma = 90$ ps and  (b) the corresponding  $\lambda_{\text{sim}}(\epsilon)$ for $\Delta \epsilon= 5$ ps with dashed line $\lambda_{\text{sim}} /T = \text{ln}(2)$ and labeling of the microscopic, structured-mesoscopic, and macroscopic features of the SDLE. (c)  $\lambda_{\text{sim}}(\epsilon)$ computed from the model with  $\sigma = 45$ ps and (d)  $\sigma = 180$ ps.}
\label{fig:simulations}
\end{figure}

The simulated return map is shown in Fig. \ref{fig:simulations}a with a 1D structure similar to a tent map with slopes $\sim \pm 2$. The probability density of the map is also non-uniform, where clustering occurs at evenly-spaced intervals. This differs from the experimental density because, in the model, we can guarantee that $\tau_n = L\tau_1$, for integer $L$. Even though there are an integer number of logic gates in the experimental $\tau_n$, heterogeneities in gate delays due to physical effects and FPGA routing cause $\tau_n = L \tau_1 \pm \epsilon_\tau$, where $\epsilon_\tau$ is a cumulative timing difference. We implement a timing difference in the model (not shown) and note that clustering in the return map changes with $\epsilon_\tau$.

Next, we calculate the simulated SDLE $\lambda_{\text{sim}}(\epsilon)$ for the return map in Fig. \ref{fig:simulations}a. We acknowledge that there are different methods for the treatment of noise when computing Lyapunov exponents in simulated chaotic systems with noise \cite{Boffetta2002}. In our simulations, we compute iterations of the model with additive noise at each iteration and search for neighboring points. This method is equivalent to specifying nearby initial conditions and using different realizations of noise along each trajectory. Furthermore, this method allows us to emulate the process of collecting experimental data and use the same algorithm for finding the SDLE that is applied in the experiment. 

The resulting simulated SDLE is plotted in Fig. \ref{fig:simulations}b. In the figure, the SDLE increases rapidly as $\epsilon$ goes to zero, which is similar to the observed experimental SDLE. For $\epsilon < \tau_1$, the divergence of neighboring points is dominated by a stochastic, uncorrelated process provided by the additive Gaussian noise. In addition, Fig. \ref{fig:simulations}b demonstrates that $\lambda_{\text{sim}}(\epsilon)$ contains mesoscopic features $\sim O(\tau_1$) that oscillate about the average divergence rate of the macroscopic chaos.

\section{Discussion and Concluding Remarks}

Similar to the experiment, the oscillatory structure in the simulated SDLE has a frequency of $\sim 1/(2 \tau_1)$ and not $1/\tau_1$. Based on the model's construction, we attribute the $1/(2 \tau_1)$ frequency to grammar restrictions in the dynamics that are caused by symmetric gaps on either side of the map $M$ due to the discontinuities in $g$. Though the symmetry of these gaps is not visible in the experimental map, it is inherent in its design and we hypothesize that, on average, similar grammar restrictions occur in the experimental dynamics. Thus, the results from the simple model are quantitatively similar to those from the experimental Boolean system, where more agreement can likely be achieved by introducing individual gate-delay heterogeneities, using a macroscopic slope $m = 1.95$, and exploring nonlinear functions for $h$.

Furthermore, we note that the structures in the simulated SLDE become more (less) pronounced for lower (higher) levels of noise, and for sufficiently high noise levels, the structures are no longer detectable. As examples, in  Figs. \ref{fig:simulations}c-d, we include the results from simulations of the model with noise levels $\sigma = 45$ ps and $\sigma = 180$ ps. For $\sigma = 45$ ps, the amplitude of the oscillatory structure grows while its frequency remains fixed. For $\sigma = 180$ ps, the noise is on a comparable scaling to the mesoscopic $\tau_1$, which removes the structure in the SDLE. This suggests that some physical chaotic systems may have underlying fine-scaled features that do not play a role in the divergence of trajectories because they are on the same scale or smaller than the microscopic noise levels. In addition, as shown in Figs. \ref{fig:simulations}b-d, the amplitude of the noise influences the size of the microscopic region in the SDLE; larger noise amplitudes extend the microscopic region to larger epsilon \cite{Cencini2000,Gao2006}.

Lastly, introducing irregularly-spaced discontinuities in the model (not just at $\tau_1$ but at $\tau_1$, $2\tau_1$, etc.) also blurs these structures (not shown) and suggests that heterogeneities may be a mechanism that removes structures in the SDLE. In our experiment, the mesoscopic spacings between discontinuities can be tuned by moving the AND-gate inputs in Fig. \ref{fig:Stretching}a, and thus our 1D Boolean chaotic system is a good candidate to begin exploring these concepts.

In summary, we present an experimental chaotic system with a macroscopic 1D return map, microscopic noise, and mesoscopic discontinuities that are regularly spaced. We demonstrate these three scalings in a SDLE and observe that the mesoscopic features create an oscillatory structure in the SDLE. Using a physically-motivated, simple model, we reproduce a similarly structured scale-dependent Lyapunov exponent and note its effects at different noise levels. 

S.D.C acknowledges the financial support of the US Army Aviation and Missile Research Development and Engineering Center and is thankful for discussions with Dr. Ned Corron and Dr. Jonathan Blakely. S.D.C. also thanks Dr. David Rosin his procedures in Ref. \cite{RosinThesis2014}.

%
\end{document}